# Confining light wave by using non-Euclidean transformation optics

**Fei Sun** [1, 2]

1 Centre for Optical and Electromagnetic Research, State Key Laboratory of Modern Optical Instrumentation, Zhejiang University; Joint Research Centre of Photonics of the Royal Institute of Technology (Sweden) and Zhejiang University, East Building #5,Zijingang Campus, Zhejiang University(ZJU), Hangzhou 310058, China
2 Department of Electromagnetic Engineering, School of Electrical Engineering, Royal Institute of Technology (KTH), S-100 44 Stockholm, Sweden

**Abstract:** A new way to understand why some inhomogeneous dielectric medium can be used for confining the light wave has been given by using non-Euclidean transformation optics. We show that special inhomogeneous dielectric medium, which corresponds to a dip in the reference space, can be used for focusing a plane wave. If we set an absorber at the appropriate position of this medium, we can highly confine the light on the subwavelength scale.

Confining the light wave has many important applications, for instance, high efficient light harvest for solar cells [1,2], super-resolution imaging (e.g., Scanning near field optical microscope [3]), optical nano-antennas [4], nano-manufacturing and etc. However due to the diffraction limit [5], it is hard to concentrate light into the subwavelegth scale. Many various methods have been proposed to beat the diffraction limit and confine the light wave to the subwavelength scale. Many of them are based on plasmonic devices which rely on the optical property of metals [6-7]. Transformation optic has also been unitized, under the quasi-static approximation, to design plasmonic devices for controlling of light on the subwavelength scale [8]. Besides, we can also confine the light to the subwavelength scale by using all-dielectric material (e.g., photonic jets [9, 10]). In this paper, we propose a method to design a kind of inhomogeneous dielectric medium which can be used for confining light. If suitable absorbers are introduced at the appropriate position in the medium, we can highly confine the light to the subwavelength scale. Our method is based on the transformation optics (TO) [11-14] and the non-Euclidean reference space.

For simplicity, we only consider the two dimensional (2D) case in this paper. We can imagine when a plane wave propagates on a 2D plane: its amplitude should be uniform everywhere (no scattering and no concentration). However, the situation is totally different when light propagates on a non-Euclidean space: if a light wave propagates on a 2D non-Euclidean surface (e.g., a plane with a pit on it), it may naturally converge or concentrate (e.g., at the pit on the plane). Let consider a simple geometry, a plane with a spherical pit at its center (the radius of this half sphere is *a*), which is the reference space here. If a light wave propagates on this 2D surface, the light will naturally be trapped to the pit. By mapping this configuration into a plane, we will obtain a circular inhomogeneous medium with radius *a* filled in the free space. The specific transformation can be treated as two parts: if $\rho'\geq 1$, we use identical transform (the reference plane outside the pit will be identically projected onto the 2D physical plane). It means it is still the free space in this region on the physical plane. If $\rho'<1$, we use the stereographic projection [15] to project a spherical pit onto the physical plane. In the following, we will derive the material parameters in this circular region by using TO. The stereographic projection can be written as:

$$x' = \cos\varphi \cot\left(\frac{\theta}{2}\right), \quad y' = \sin\varphi \cot\left(\frac{\theta}{2}\right), \quad z' = r \qquad (1)$$

where $r$, $\theta$ and $\varphi$ are in the spherical coordinate system of the reference space. $x'$, $y'$ and $z'$ are in Cartesian coordinate system of the physical space. We should note the quantities with and without quotation mark indicates quantities are in physical space and reference space respectively. We have assumed the unit of the length is *a* in the derivation of the above (*a* is the radius of the spherical pit in reference space and also the radius of the circle in physical space). It means all physical quantities with unit of the length have been normalized by *a* in this paper. We assume it is the free space on the reference Non-Euclidean surface. The material in the physical space corresponding to the pit in the reference space can be calculated by using the formula of TO [14]:

$$\bar{\bar{\varepsilon}}'/\varepsilon_0 = \bar{\bar{\mu}}'/\mu_0 = -diag\left(1,1,\left(\frac{2}{1+\rho'^2}\right)^2\right) \qquad (2)$$

Eq. (2) is the material filled in the circular region with a radius *a*. The wave equation for the TE mode (the electric field is normal to the propagation plane) in a 2D plane can be written as:

$$\frac{\partial}{\partial x'}\left(\frac{1}{\mu_y{'}}\frac{\partial E_z{'}}{\partial x'}\right)+\frac{\partial}{\partial y'}\left(\frac{1}{\mu_x{'}}\frac{\partial E_z{'}}{\partial y'}\right)+\omega^2\varepsilon_z{'}E_z{'}=0 \quad (3)$$

As we can see from Eq. (2), $\mu_x=\mu_y=-1$ are constant in the circular region, thus the wave equation (3) can be rewritten as:

$$\left(\frac{\partial^2}{\partial x'^2}+\frac{\partial^2}{\partial y'^2}\right)E_z{'}+\omega^2(-\varepsilon_z{'})E_z{'}=0 \quad (4)$$

It means that the refraction index in this circle region with radius $a$ should satisfy:

$$n'=\sqrt{-\varepsilon_z{'}}=\frac{2}{1+\rho'^2},\rho'<1 \quad (5)$$

The medium described by Eq. (5) is filled in the free space ($n'=1$ for $\rho'\geq 1$) in physical space, which (the whole configuration) behaves equivalently like a non-Euclidean 2D surface (a plane with a spherical pit in its center in reference space) for the TE polarization wave. The medium inside the circle is the same with the so-called Maxwell's fish eye lens (MFL) [16]. In recent years, more and more attentions have been focused on the resolution of MFL [17-19]. In this paper, our attention is focused on the subwavelength focusing with the help of the MFL. We should note that our design method is a general method. The reference space can be a plane with a pit of other shapes (not necessary a spherical pit). If we choose some other pits of various shapes, we will obtain other inhomogeneous medium but not the MFL. We will further study pits of various shapes and different illumination for focusing applications in the following work.

We use the finite element method (FEM) to verify that the medium described by Eq. (5) can be used for confining the light wave. When the wavelength of an incident plane wave is $\lambda_0=2.5a$, the light can be confined in the inhomogeneous medium (see Fig. 1). However the concentrated spot is still diffraction limited in this case.

When a light wave propagates on a plane with a spherical pit at its center, it will naturally be trapped to the pit. However if we do not set an absorber in the pit to pick up the light, the light wave will slip away from the pit. That is the reason why the concentrated spot in Fig. 1 is diffraction limited. In practical, we can easily put a photonic detector or an absorber at the center of the medium described by Eq. (5). It will capture the light wave and help to confine the light wave on the subwavelength scale. We also use FEM to verify this. As coaxial cables have been used as drains in MFL for imaging applications [20], we also use a coaxial cable with a core diameter of $0.01a$ and an outer diameter of $0.02a$ as the absorber here for focusing applications. If we introduce a coaxial cable, we have to make a three dimensional (3D) simulation. However the medium designed in this paper is 2D case. A cylindrical medium which is uniform in $z$ direction and inhomogeneous in $x$-$y$ plane (described by Eq. (5)) is put into two parallel perfect electric conductor (PEC) plates. The separation between two PEC plates is extremely small ($0.05a$) to mimic a 2D space between them. The coaxial cable is inserted from the bottom PEC plate $0.03a$ exposed into the waveguide structure. We put the coaxial cable at the point where electric field is maxima but not the geometrical center of the medium. The simulation result is shown in the Fig. 2. By introducing an absorber at the suitable position of designed medium, the light wave (wavelength can be much larger than the size of the medium) can be highly confined on the subwavelength scale.

In conclusion, a new way to understand why some inhomogeneous medium can be used to confine light wave has been given by using non-Euclidean transformation optics. With the help of some absorbers, we can further confine the light wave to subwavelength scale in those inhomogeneous mediums.

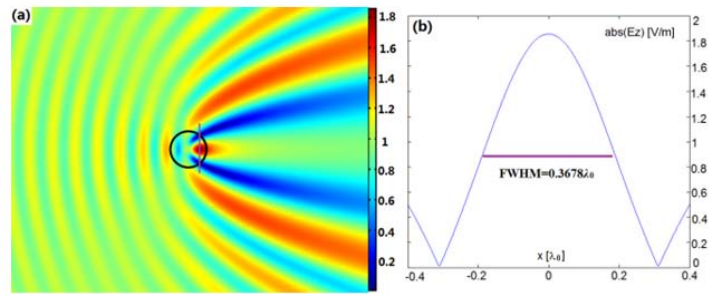

Fig. 1. The absolute value of electric field distribution: (a) in the 2D plane and (b) on the purple cut line in (a). A TE polarized plane wave with unit amplitude incidents to the medium from left to right. The wavelength of the incident wave is $\lambda_0=2.5a$. It is the free space outside the black circle and the medium described by Eq. (5) inside the black circle.

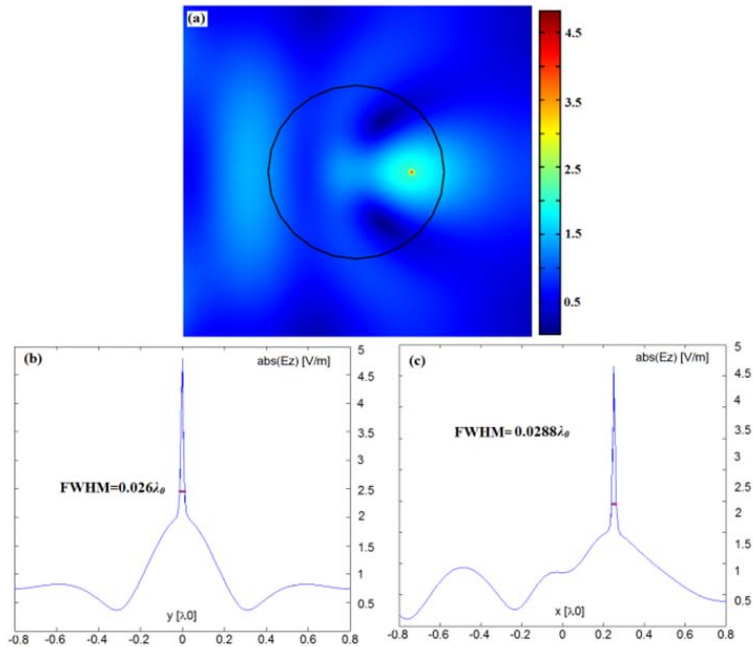

Fig. 2. The absolute value of electric field distribution when we insert a coaxial cable as an absorber at the point with maxima field in Fig. 1. The wavelength of the incident plane wave is $\lambda_0=2.5a$. The medium is put into a two parallel PEC plate separated by $0.05a$. The coaxial cable (inserted from the bottom PEC plate $0.03a$ exposed into the waveguide structure) is used as an absorber here. (a) The absolute value of electric field distribution on the 2D plane ($0.015a$ above the coaxial cable). (b) and (c) the cut line along the maxima field point of (a) in $y$ and $x$ direction respectively.